# Obstruction-Driven Parity Inversion for Enhanced Optical Absorption in Hexagonal Transition Metal Dichalcogenides

**Authors:** Seungil Baek[1], Jun Jung[1], and Yong-Hyun Kim[1,2,*]

[1]Department of Physics, Korea Advanced Institute of Science and Technology (KAIST), Daejeon 34141, Republic of Korea

[2]School of Physics, Institute of Science, Suranaree University of Technology, Nakhon Ratchasima 30000, Thailand

*Correspondence to: yong.hyun.kim@kaist.ac.kr

The optical selection rule states that opposite parity between the valence and conduction bands is required for optical absorption to occur. However, monolayer hexagonal transition metal dichalcogenides (h-TMDs) such as $MoS_2$ exhibit pronounced optical absorption despite their nominally dipole-forbidden *d-d* transitions. In this Letter, we elucidate a parity inversion mechanism through which obstruction-driven band inversion promotes dipole-allowed optical transitions near the band edge in monolayer h-TMDs. By comparing trivial and obstructed atomic limit phases, we show that intersite interactions between hybridized *d* orbitals induce parity inversion. Our results provide a novel approach to tuning optical properties through parity control, bridging the gap between topology and light-matter interaction.

The optical selection rule requires that the initial and final states should possess different parity for an optical transition to occur [1]. For instance, transition metal oxides exhibit minute optical absorption due to their *d-d* transition character, only by virtue of even-odd parity mixing [2–5]. In contrast, monolayer hexagonal transition metal dichalcogenides (h-TMDs), despite being only a single atom thick, exhibit large optical absorption, whose sunlight absorption is comparable with few-nm-thick Si or GaAs [6–8]. This exceptional absorption has stimulated extensive studies such as near-perfect light absorption [9–11], photovoltaics [12,13], and photocatalysis [14–16]. Band nesting, referring to nearly parallel conduction and valence bands, has been theoretically established as the underlying origin of the strong optical absorption [17,18]. However, considering that the band edges of h-TMDs are derived from transition metal (TM) *d*-orbitals, it remains unclear what mechanism relaxes the optical selection rule and enables a finite dipole matrix element.

Traditionally, crystal field splitting was thought to be the origin of the band gap in h-TMDs [19–21]. However, the crystal field preserves atomic orbital characters, which is contradictory to the presence of the nontrivial band inversion of h-TMDs at $\pm K$ point. Moreover, the building blocks of band edges in h-TMDs, $d_{z^2}$, $d_{xy}$, and $d_{x^2-y^2}$ orbitals, do not allow optical transitions in between themselves. Recently, the hidden breathing kagome lattice (BKL) structure has been proposed as a fundamental framework for understanding the electronic structure of h-TMDs [22,23]. In h-TMDs, $sp^2$-like hybridization of three *d* orbitals that constitute band edges leads to hybrid *d* orbitals, which are in-plane $d_{z^2}$ orbitals directing metal-chalcogen bond directions. Along with the intersite interaction, the crystal field acting as intra-cell hopping, or *on-site hopping*, between these hybrid *d* orbitals establishes a breathing kagome lattice (BKL). As with the Su-Schrieffer-Heeger (SSH) model [24], two topologically distinct phases could be obtained in BKL by the competition between intersite and on-site hoppings [23,25–28]. The dominance of on-site hoping results in trivial atomic insulators (AL), where the Wannier center (WC) coincides with the atomic site. On the other hand, the stronger intersite hopping results in an obstructed atomic limit (OAL) insulator, where WC is located away from atomic sites. Although both are classified to possess trivial band topology, OAL and AL cannot be adiabatically connected without closing the band gap [26–29]. As h-TMDs have stronger intersite hopping than on-site hopping, they naturally possess the OAL nature, characterized by off-atom WC and band inversion [23,30,31].

Recent studies suggest a direct relationship between topology and optical absorption through quantum geometry [32]. In particular, Onish *et al*. claimed that the elevated quantum metric, the real symmetric part of the quantum geometric tensor, near the topological band inversion is the origin of large optical absorption in topological insulators [33]. Certain findings on Dirac Hamiltonians motivate further exploration of this possibility [33–35]. On the other hand, BKL and h-TMDs exhibit band inversion without hosting nontrivial band topology. This naturally raises the question of whether band inversion in such systems can likewise enhance optical conductivity. It is also elusive how the band inversion could strengthen the optical response. Moreover, the OAL insulators benefit from their comparably larger band gap than conventional topological insulators, a feature that emphasizes the importance of studying their optical responses.

In this Letter, we analyze the obstruction-driven parity inversion mechanism in monolayer h-TMDs, which enables dipole-allowed optical transitions. By comparing two tight-binding Hamiltonians that are OAL and AL phases, we show that the OAL phase with band inversion indeed features enhanced optical absorption. Since atomic orbitals are no longer an adequate basis for OAL insulators, we interpret h-TMDs in terms of molecular orbitals composed of hybrid *d* orbitals. Our analysis reveals that bonding interactions between these hybrid *d* orbitals give rise to finite optical absorption, whereas the crystal field effect plays no role. These findings provide an orbital-level mechanism for understanding the strong optical absorption in h-TMDs, revealing orbital hybridization as the microscopic bridge between topology and optical responses.

Density functional theory (DFT) calculations were performed using Vienna ab initio simulation package (VASP) with the projector-augmented wave (PAW) pseudopotentials [36,37]. Perdew-Burke-Ernzerhof (PBE) exchange-correlation functional was used [38]. The energy cutoff of the plane-wave basis was set to 400 eV, and Γ-centered 24 × 24 × 1 *k*-points were used. Maximally localized Wannier functions (MLWF) and MLWF-interpolated band structures were obtained using Wannier90 [39]. We constructed three-band tight-binding (TB) Hamiltonians with up to third nearest-neighbor hoppings taken into account, based on the TM *d*-orbitals [23,40]. These Hamiltonians were fitted into the Wannier-interpolated band structures with root mean square errors smaller than 5 meV. Adjusting TB parameters results in two distinct electronic structures, with and without band inversion features, yet their band dispersions are nearly identical. Atomic and electronic structures were visualized with VESTA [41].

The Kubo-Greenwood formula for the non-interacting electron systems was used to calculate optical conductivity [42,43]

$$\sigma_{\alpha\beta}(\omega) = \frac{e^2}{\hbar} \int \frac{d^2k}{(2\pi)^2} \sum_{n,m} (f_{n\mathbf{k}} - f_{m\mathbf{k}}) \frac{i(\epsilon_{m\mathbf{k}} - \epsilon_{n\mathbf{k}}) A^{\alpha}_{nm\mathbf{k}} A^{\beta}_{mn\mathbf{k}}}{(\epsilon_{n\mathbf{k}} - \epsilon_{m\mathbf{k}}) + \hbar\omega + i\delta}$$

where $e$ is the fundamental charge, $\epsilon_{n\mathbf{k}}$ is the eigenvalue at $\mathbf{k}$, $f_{n\mathbf{k}} = \left[1 + \exp\left(\frac{\epsilon_{n\mathbf{k}} - \epsilon_F}{k_B T}\right)\right]^{-1}$ is the Fermi-Dirac distribution of the $n$-th band at $\mathbf{k}$ with the Fermi level $\epsilon_F$ and the temperature $T$, $A^{\alpha}_{nm\mathbf{k}} = \langle u_{m\mathbf{k}} | i\partial_{\alpha} | u_{n\mathbf{k}} \rangle$ is the interband Berry connection, and $|u_{n\mathbf{k}}\rangle$ is the periodic part of the Bloch wave function. Here, the eigenvalues and interband Berry connections are computed from three-band TB models with 4000 × 4000 × 1 $k$-points sampled. The Lorentzian broadening of $\delta$ = 1 meV was used for the calculations. To study the momentum-resolved optical behaviors, the transition dipole moment (TDM) from the initial state $a$ to the final state $b$ was calculated as

$$\mathbf{P}_{a \to b} = \langle u_{b\mathbf{k}} | \mathbf{r} | u_{a\mathbf{k}} \rangle = \frac{i}{(\epsilon_{b\mathbf{k}} - \epsilon_{a\mathbf{k}})} \langle u_{b\mathbf{k}} | \nabla_{\mathbf{k}} H_{\mathbf{k}} | u_{a\mathbf{k}} \rangle$$

where the $k$-derivative of the tight-binding Hamiltonian $H_{\mathbf{k}}$ was obtained by finite differences. The squared magnitudes $|\mathbf{P}_{a \to b}|^2$ are presented in Debye$^2$.

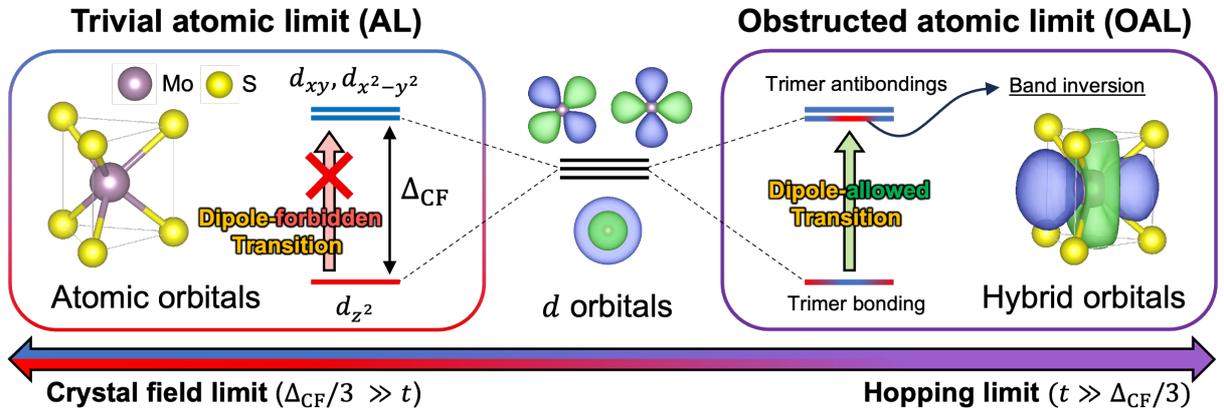

FIG. 1. Schematic illustration of dipole-allowed and forbidden transitions. In the crystal field limit, the atomic orbital character is preserved with trivial atomic limit character, resulting in forbidden optical transitions. On the other hand, in the hopping limit, intersite hopping induces

parity inversion with obstructed atomic limit character, which enables pronounced optical transitions.

The atomic structure of h-TMDs is trigonal prismatic, where the TM atom (Mo or W) is surrounded by six chalcogen atoms (S, Se, or Te). The band edge characters of h-TMDs are governed by three *d*-orbitals of TM, which are $d_{z^2}$, $d_{xy}$, and $d_{x^2-y^2}$. The crystal field lifts the degeneracy of the *d*-orbitals, splitting them into $d_{z^2}$ and $\{d_{xy}, d_{x^2-y^2}\}$ by an energy $\Delta_{\text{CF}}$, as shown in Fig. 1. The *sp*$^2$-like hybridization of $d_{z^2}$, $d_{xy}$, and $d_{x^2-y^2}$ orbitals produce three *lie-down* $d_{z^2}$ orbitals [23], heading along TM-chalcogen bond direction. These orbitals are called hybrid *d* orbitals, which provide a single dominant hopping channel ($t_1$) between nearest neighbors. Furthermore, an originally crystal-field-induced splitting ($\Delta_{\text{CF}}$) now transforms into the hoppings between hybrid *d* orbitals ($t_2 = \Delta_{\text{CF}}/3$). As these hybrid *d* orbitals are all located at the same TM site, these hoppings are called on-site hoppings. A former study demonstrated that intersite hopping ($t_1$) and on-site hopping ($t_2$) between hybrid *d* orbitals organize an electronic breathing kagome lattice (BKL) (See Fig. S1) [23,44].

For h-TMDs, intersite hopping was dominant than on-site hopping ($t_1 > t_2$). In this case, the charge centers are located at off-atomic sites rather than at atomic sites, corresponding to the OAL insulator. They are distinct from an AL insulator, in which the charge centers coincide with atomic sites. In particular, the OAL phase cannot be adiabatically connected to AL without closing the band gap. Although categorized as topologically trivial insulators, OAL insulators are topologically distinct and cannot be adiabatically connected to AL insulators. In the case of BKL, the topological characteristics of OAL and AL could be characterized by band inversion at the K point. Likewise, h-TMDs exhibit band inversion at the K point, which suggests that the hopping strength is large enough to overcome $\Delta_{\text{CF}}$. Note that there is an ongoing debate about the nontrivial topology of BKL, particularly concerning the robustness of edge states [25–28]. However, our study focuses on the obstructed atomic limit and relevant band inversion at the K point.

As illustrated in Fig. 1, the origin of the band gap determines whether optical absorption happens or not. In the crystal field limit, each band corresponds to the atomic orbitals, sharing identical angular momentum ($l = 2$). Since band edges of h-TMDs are comprised of *d*-orbitals of TM, no dipole transitions are allowed in this case. On the other hand, intersite interaction could also introduce energy splitting in these materials. Large intersite hopping makes hybrid *d* orbitals a suitable basis in this system, and bonding interactions between these orbitals break the original parity of atomic orbitals. This mechanism relaxes the optical selection rule, allowing a dipole transition. That is, the OAL character induced by intersite hopping is the key to enhanced optical absorption. One cannot account for the optical transition by orbital character alone, as the interaction may reverse what the orbital character suggests.

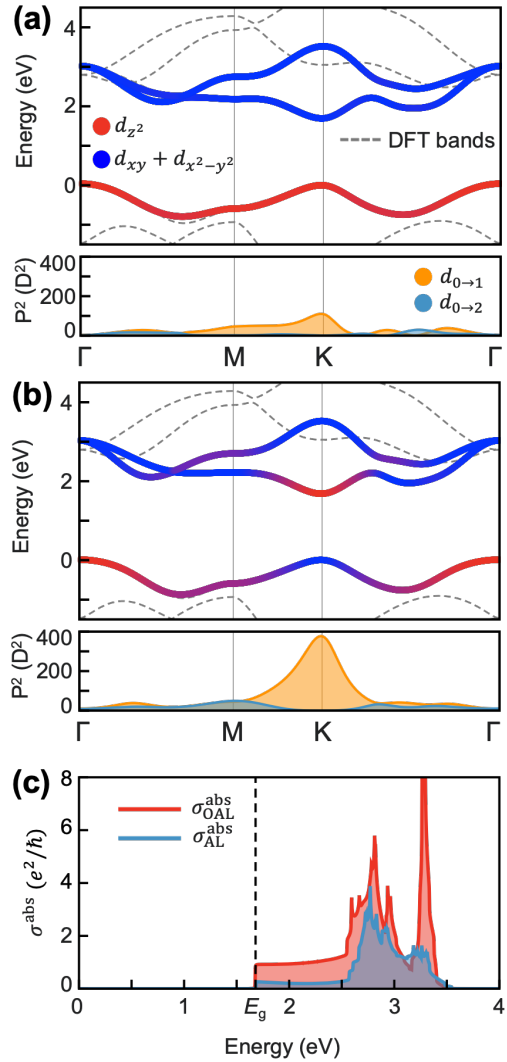

FIG. 2. Comparison of optical conductivities in trivial and topological phases of $MoS_2$. (a), (b) Electronic band structures and transition dipole moments for the trivial phase and the topological phase, respectively. Transition dipole moments are calculated from the three-band tight-binding model. The DFT band structure is shown with gray dashed lines. (c) Optical

absorption spectra of topological (OAL, red) and trivial phase (AL, blue) calculated from the three-band model. Here, the OAL phase exhibits about four times larger optical absorbance in the low-energy regime.

To confirm this idea, we constructed a three-band tight-binding (TB) model for h-TMDs, with lattice symmetries taken into account. Based on TM $d$-orbitals, this TB model accounts for the topmost valence band (VB) and the two lowermost conduction bands (CB) of h-TMDs. It was previously reported that the incorporation up to third-nearest-neighbors preserves band dispersion of h-TMDs and band inversion feature at the K point [23,40]. As a prototypical example of h-TMDs, MoS$_2$ was chosen for this study. We have fitted our TB Hamiltonians into DFT band structures, but in two distinct targeting phases. The first one is an ordinary model that depicts h-TMDs, featuring band inversion. This phase represents a hopping-dominant phase or OAL insulator. Here, the two dominant hopping parameters were $t_1 = 0.82$ eV and $t_2 = 0.41$ eV, which were a close resemblance to $t_1 = 0.89$ eV and $t_2 = 0.26$ eV obtained from MLWFs. The other one is an artificial phase without band inversion, indicating a crystal field dominant phase or an AL insulator. In contrast to the OAL phase, fitted TB parameters were $t_1 = 0.24$ eV and $t_2 = 0.89$ eV, illustrating dominance of on-site hopping (See Table. S1 for detailed TB parameters) [44]. Figs. 2(a) and 2(b) demonstrate band structures plotted from the obtained AL and OAL insulating phases, respectively. Note that they share the same band dispersion, but the OAL phase displays $d_{xy}$ and $d_{x^2-y^2}$ (blue) character at valence band maximum (VBM) and $d_{z^2}$ character (red) at conduction band minimum (CBM).

We calculated the optical absorption characteristics of h-MoS$_2$. We first compared transition dipole moments (TDM) along the $k$-space path connecting high-symmetry points. The magnitude of TDM represents the transition probability between two states at specific $k$-points. The lower figures in Figs. 2(a) and 2(b) demonstrate calculated TDM values. In the AL phase, the maximum magnitude of TDM from VBM to CBM ($d_{0\to1} = |\mathbf{P}_{0\to1}|^2$) was 94.1 Debye$^2$. However, in the OAL phase, the maximum $d_{0\to1}$ was 348.9 Debye$^2$, which is about 3.7 times greater than the AL phase. For both cases, TDM at the Γ point is negligible as symmetry constrains the eigenstate to be close to the atomic orbitals. Furthermore, we calculated the absorptive part of optical conductivity through the Kubo-Greenwood formula, which is given in Fig. 2(c). Although our results do not encompass high-energy features of h-TMDs, both phases exhibit step-function-like behavior for energies larger than the band gap, which is a typical feature in 2D systems. Moreover, as predicted by TDM, the OAL phase shows about 4 times larger optical conductivity than the AL phase in the low energy regime. Our calculation on OAL optical conductivity agrees with the one obtained from the MLWFs (See Fig. S2) [44].

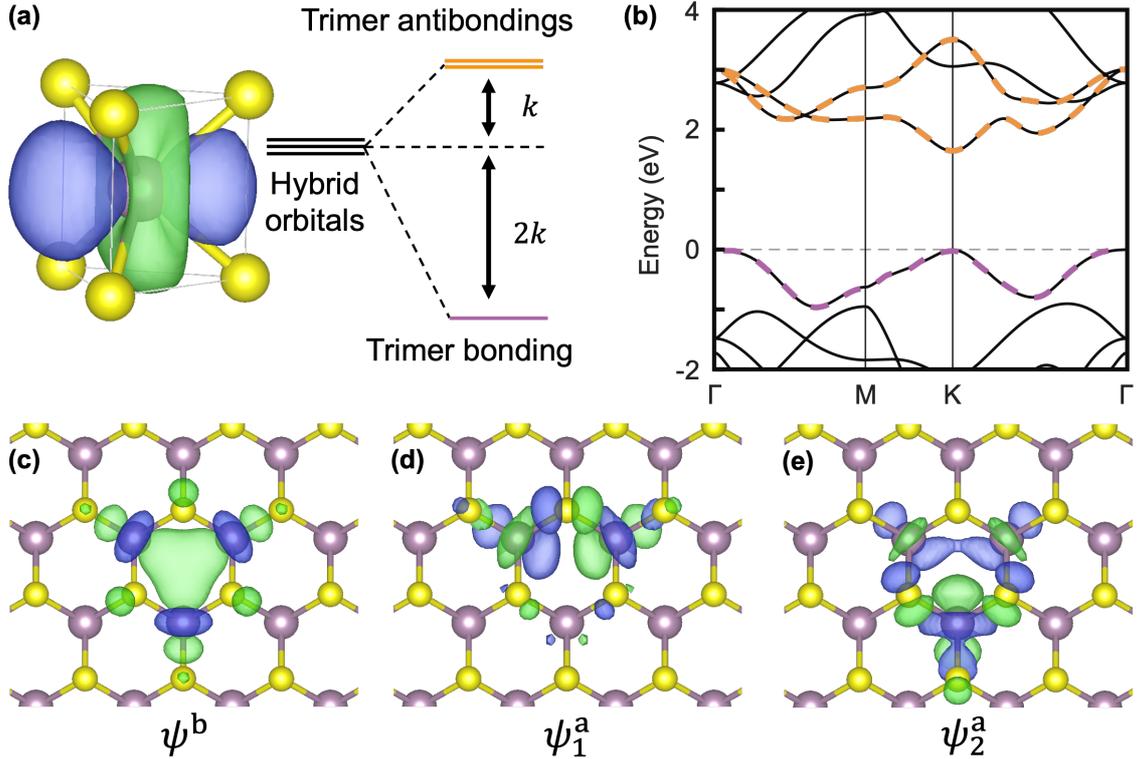

FIG. 3. Molecular orbital structure of h-MoS$_2$. (a) Intersite hopping interaction between hybrid *d* orbitals results in a trimer bonding state and doubly degenerate trimer antibonding states. (b) The electronic band structure of MoS$_2$ is shown in the black line. Wannier-interpolated bands for the valence band (purple) and the conduction bands (orange) are overlaid as dashed lines. (c) Maximally localized Wannier function of the trimer bonding state ($\psi^b$). (d), (e) Maximally localized Wannier functions of the antibonding states ($\psi_1^a, \psi_2^a$), respectively. Note that the bonding and antibonding states have different parity.

Wannier functions reveal the origin of such enhanced optical absorption in the OAL phase. Since the OAL phase does not have a charge center on the atomic sites, the basic unit of charge should be rather called a molecular orbital. Figure 3 illustrates the molecular orbital structure of h-MoS$_2$. Once we neglect the on-site interaction, the intersite interaction between three degenerate hybrid *d* orbitals results in one trimer bonding state and two degenerate trimer antibonding states, as shown in Fig. 3(a). Here, we note that the intersite interaction constitutes another origin of the band gap in h-TMDs, in addition to crystal field splitting. The resulting band gap is a hybridization gap between bonding and antibonding states. Our MLWF results confirm this idea. In Fig. 3(b), MLWF interpolation reveals trimer bonding (purple) and antibonding states (orange) that correspond to the valence and conduction bands of MoS$_2$ (black). The Wannier orbital of VB ($\psi^b$) is plotted in Fig. 3(c), whose charge center reconfirms the OAL character. In addition, Wannier orbitals of two degenerate CBs ($\psi_1^a$, $\psi_2^a$) are demonstrated in Figs. 3(d) and 3(e).

The VB Wannier orbital ($\psi^b$) is a bonding state of a neighboring three hybrid *d* orbitals, sharing a single hexagonal ring. Since these three orbitals combine with the identical phase, we can consider this trimer bonding state as an effective *s* orbital. It follows three-fold rotational symmetry ($C_3$), and is even to the *yz* plane. On the other hand, trimer antibonding states ($\psi_1^a$, $\psi_2^a$) have nodal planes along the *yz* and *xz* planes, respectively. Thus, $\psi_1^a$ and $\psi_2^a$ could be considered as effective $p_x$ and $p_y$ orbitals on a hexagonal ring, respectively. As a consequence, though comprised of TM *d*-orbitals, molecular orbitals on VB and CB now possess opposite parity, even (*gerade*) for trimer bonding state and odd (*ungerade*) for trimer antibonding states. The optical selection rule, obligating inverted parity between initial and final states, is alleviated by this parity inversion mechanism. The orbital overlap between trimer bonding and antibonding states is indeed zero, $\langle \psi^b | \psi_i^a \rangle = 0$ for *i* = 1, 2. This *obstruction-driven parity inversion* leads to the enhanced optical absorption, as we have already discussed in Fig. 2.

|  |  | TDM (D$^2$) | $\sigma_{abs}$ ($e^2/\hbar$) |  |  | TDM (D$^2$) | $\sigma_{abs}$ ($e^2/\hbar$) |
|---|---|---|---|---|---|---|---|
| MoS$_2$ | OAL | 376.9 | 0.462 | WS$_2$ | OAL | 469.9 | 0.436 |
|  | AL | 110.3 | 0.135 |  | AL | 133.1 | 0.134 |
| MoSe$_2$ | OAL | 362.3 | 0.436 | WSe$_2$ | OAL | 472.7 | 0.404 |
|  | AL | 107.1 | 0.125 |  | AL | 152.2 | 0.132 |
| MoTe$_2$ | OAL | 410.1 | 0.336 | WTe$_2$ | OAL | 635.6 | 0.310 |
|  | AL | 254.6 | 0.173 |  | AL | 393.0 | 0.174 |

Table 1. Comparison of optical properties of OAL and AL phases of h-TMDs. TDM at the K point and $\sigma_{abs}$ at the band gap are given in Debye$^2$ and $e^2/\hbar$, respectively.

The obstruction-driven parity inversion mechanism is universal for h-TMDs, as they share the hidden BKL structure [23]. We performed the same calculations for the family of h-TMDs (MX$_2$), with M = Mo or W and X = S, Se, or Te. For all cases, the three-band TB model well described the electronic structures of h-TMDs. The TB Hamiltonians were constructed for h-TMDs in both AL and OAL phases, where their TB parameters and Wannier centers are given in the Table. S1-S4 [44,45]. As in MoS$_2$, OAL phases exhibited the dominance of intersite hopping, while AL phases had comparably larger crystal field splitting. The maximum TDM and optical absorption at the band gap for each material and phase are shown in the Table. 1 and Fig. S3 [44]. In particular, TDM and optical absorption were both notably higher in the OAL phase than AL phase for all materials. Moreover, the MLWFs of h-TMDs share the identical feature, the trimer bonding and antibonding characters (see Fig. S4) [44]. Accordingly, we establish that the obstruction-driven parity inversion mechanism is generally applicable to h-TMDs.

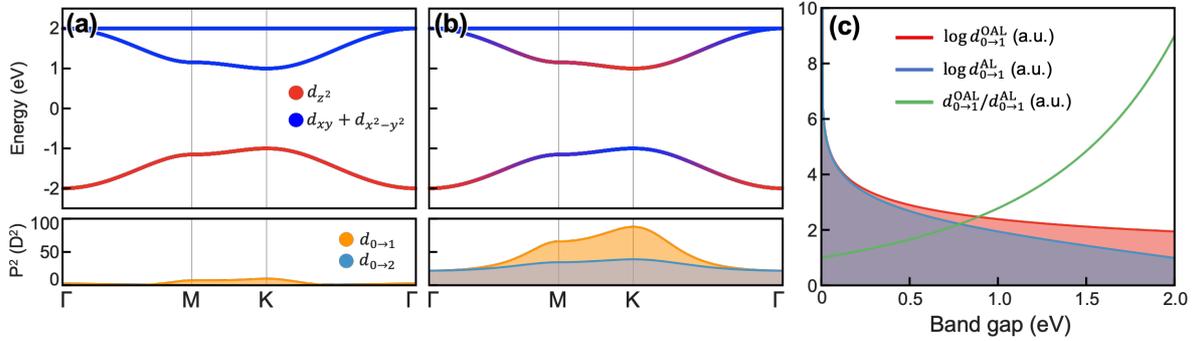

FIG. 4. Comparison of the simplified model systems in the crystal field and hopping limits. Electronic band structures and transition dipole moments (P) are presented. (a) Atomic limit phase. (b) Obstructed atomic limit phase. Although sharing identical band dispersion, band-inverted phase exhibit enhanced transition dipole moments. (c) The maximum transition dipole moments as a function of band gap for different breathing strengths. The band dispersions remain identical when the band gap is the same.

To investigate the interplay between intersite hopping and crystal field, we compared a two-parameter model where only $t_1$ and $t_2$ are nonzero. Note that this model is equivalent to the BKL with the nearest neighbor interaction. Figure 4 compares simplified h-TMD models. The band structures and TDMs of AL phase ($t_1 = -1/3$ eV and $t_2 = 1$ eV) and OAL phase ($t_1 = -1$ eV and $t_2 = 1/3$ eV) are given in Figs. 4(a) and 4(b), respectively. Although they share the identical band dispersion, their orbital composition varies, with and without band inversion. The difference originates from the relative dominance of intersite hopping versus the crystal field. In like manner with Fig. 2, relatively preserved atomic orbital characters in AL phase lead to small TDM (~9.7 Debye$^2$) in Fig. 4(a), while obstruction-driven parity inversion in OAL phase reinforces TDM up to ~87.7 Debye$^2$ in Fig. 4(b). In between Figs. 4(a) and 4(b), the balanced values of $t_1$ and $t_2$ yield the ideal kagome lattice ($t_1 = -2/3$ eV and $t_2 = 2/3$ eV) as illustrated in Fig. S5 [44]. In this case, the disappearance of the band gap at the K point leads to the divergence of TDM. By linearly interpolating TB parameters of AL and OAL phases, we traced the evolution of the maximum TDM between the two atomic limits. Figure 4(c) shows the maximum TDM for the same band gap but with different phases. Here, the gamma point eigenvalues and the uppermost bands remain unchanged throughout the evolution (See Fig. S5) [44]. We confirmed that the Wannier center stays fixed throughout the AL phase and up to the ideal kagome limit, whereas in the OAL phase it shifts to a different position. Although their band structures are symmetric with respect to the ideal kagome structure, the maximum TDM is clearly greater in the OAL phase than in the AL phase. Note that the TDM ratio of OAL over AL, indicated in the green line, increases as the band gap increases, which agrees with our findings that tellurides ($MoTe_2$, $WTe_2$) exhibit the smallest value of TDM ratio.

Furthermore, we could expedite the limit where either the crystal field or intersite hopping acts alone (Fig. S6) [44]. By choosing $t_1 = 0$ eV and $t_2 = 1$ eV for the crystal field limit and $t_1 = -1$ eV and $t_2 = 0$ eV for the hopping limit, we were able to compare the AL and OAL phases without a momentum-dependent effect. Given only nearest-neighbor interaction, each hexagonal ring becomes an independent counting unit in the hopping limit, and therefore, the bands are dispersionless. Likewise, the absence of intersite interaction in the crystal field limit brings flat bands, having the same eigenvalues as the hopping limit. However, the hopping limit displays nontrivial orbital structure, the band inversion at the K point. In the valence band at the $\Gamma$ point, the identical phase of three hybrid $d$ orbitals results in $d_{z^2}$ orbital of the Mo atom. On the other hand, the relative phase difference ($e^{2\pi i/3}$) between Mo atoms leads to the superposition of $d_{xy}$ and $d_{x^2-y^2}$ orbitals on Mo (Fig. S7) [44]. The TDM values were constant along *k*-space, which were ~38.7 Debye$^2$ for the hopping limit and 0 Debye$^2$ for the crystal field limit. We conclude that the crystal field does not contribute to optical transitions, while intersite hopping enables it by inducing parity inversion of molecular orbitals.

In summary, we have demonstrated that the pronounced optical absorption in monolayer h-TMDs originates from an obstruction-driven parity inversion mechanism. By constructing TB Hamiltonians for OAL and AL phases, we identified that the OAL phase with band inversion exhibits strongly enhanced optical responses. Our MLWF analysis revealed that the valence and conduction bands correspond to the bonding and antibonding states of hybrid *d* orbitals, respectively, which enable dipole-allowed transitions. Our systematic comparison across the family of h-TMDs confirmed that this mechanism is universal, establishing that the dominance of intersite hopping over crystal field splitting is the microscopic origin of strong light-matter interaction in these materials. This highlights that it is the type of interaction that governs the optical transition, rather than the orbital character itself. As our proposed mechanism heavily relies on the band inversion feature of the OAL phase, our understanding could further expand to the topological insulating systems. We hope our study to link the gap between topology and optical response via orbital hybridization, proposing parity inversion as a route to tailor optical properties.

## ACKNOWLEDGEMENTS

We thank H.-H. Nahm for fruitful discussions. This work is supported by the National Research Foundation of Korea (NRF) Grant (2019M3D1A1078302).

S.I.B. and Y.-H.K. developed the theory, performed the calculations, and wrote the manuscript. J.J. contributed to the early theoretical formulation. All authors commented on the manuscript.

# REFERENCES

[1] O. Laporte and W. F. Meggers, Some Rules of Spectral Structure, J. Opt. Soc. Am. **11**, 459 (1925).

[2] J. H. Van. Vleck, The Puzzle of Rare-earth Spectra in Solids, J. Phys. Chem. **41**, 67 (1937).

[3] R. Newman and R. M. Chrenko, Optical Properties of Nickel Oxide, Phys. Rev. **114**, 1507 (1959).

[4] G. W. Pratt and R. Coelho, Optical Absorption of CoO and MnO above and below the Néel Temperature, Phys. Rev. **116**, 281 (1959).

[5] R. J. Powell and W. E. Spicer, Optical properties of NiO and CoO, Phys. Rev. B **2**, 2182 (1970).

[6] M. Bernardi, M. Palummo, and J. C. Grossman, Extraordinary Sunlight Absorption and One Nanometer Thick Photovoltaics Using Two-Dimensional Monolayer Materials, Nano Lett. **13**, 3664 (2013).

[7] L. Britnell, R. M. Ribeiro, A. Eckmann, R. Jalil, B. D. Belle, A. Mishchenko, Y.-J. Kim, R. V. Gorbachev, T. Georgiou, S. V. Morozov, A. N. Grigorenko, A. K. Geim, C. Casiraghi, A. H. Castro Neto, K. S. Novoselov, Strong Light-Matter Interactions in Heterostructures of Atomically Thin Films, Science **340**, 1311 (2013).

[8] M. Bernardi, C. Ataca, M. Palummo, and J. C. Grossman, Optical and Electronic Properties of Two-Dimensional Layered Materials, Nanophotonics **6**, 479 (2017).

[9] S. Gupta, S. N. Shirodkar, A. Kutana, and B. I. Yakobson, In Pursuit of 2D Materials for Maximum Optical Response, ACS Nano **12**, 10880 (2018).

[10] J. Horng, E. W. Martin, Y.-H. Chou, E. Courtade, T.-C. Chang, C.-Y. Hsu, M.-H. Wentzel, H. G. Ruth, T.-C. Lu, S. T. Cundiff, F. Wang, and H. Deng, Perfect Absorption by an Atomically Thin Crystal, Phys. Rev. Appl. **14**, 024009 (2020).

[11] S. Lee, D. Seo, S. H. Park, N. Izquierdo, E. H. Lee, R. Younas, G. Zhou, M. Palei, A. J. Hoffman, M. S. Jang, C. L. Hinkle, S. J. Koester, and T. Low, Achieving near-perfect light absorption in atomically thin transition metal dichalcogenides through band nesting, Nat. Commun. **14**, 3889 (2023).

[12] Y. J. Zhang, T. Ideue, M. Onga, F. Qin, R. Suzuki, A. Zak, R. Tenne, J. H. Smet, and Y. Iwasa, Enhanced intrinsic photovoltaic effect in tungsten disulfide nanotubes, Nature **570**, 349 (2019).

[13] C. M. Went, J. Wong, P. R. Jahelka, M. Kelzenberg, S. Biswas, M. S. Hunt, A. Carbone, and H. A. Atwater, A new metal transfer process for van der Waals contacts to vertical Schottky-junction transition metal dichalcogenide photovoltaics, Sci. Adv. **5**, eaax6061 (2019).

[14] Q. Lu, Y. Yu, Q. Ma, B. Chen, and H. Zhang, 2D Transition-Metal-Dichalcogenide-Nanosheet-Based Composites for Photocatalytic and Electrocatalytic Hydrogen Evolution Reactions, Adv. Mater. **28**, 1917 (2016).

[15] Y. Fan, J. Wang, and M. Zhao, Spontaneous full photocatalytic water splitting on 2D $MoSe_2/SnSe_2$ and $WSe_2/SnSe_2$ vdW heterostructures, Nanoscale **11**, 14836 (2019).

[16] R. Yang, Y. Fan, Y. Zhang, L. Mei, R. Zhu, J. Qin, J. Hu, Z. Chen, Y. H. Ng, D. Voiry, S. Li, Q. Lu, Q. Wang, J. C. Yu, and Z. Zeng, 2D Transition Metal Dichalcogenides for Photocatalysis, Angew. Chem. Int. Ed. **62**, e202218016 (2023).

[17] A. Carvalho, R. M. Ribeiro, and A. H. Castro Neto, Band nesting and the optical response of two-dimensional semiconducting transition metal dichalcogenides, Phys. Rev. B **88**, 115205 (2013).

[18] Y. Li, A. Chernikov, X. Zhang, A. Rigosi, H. M. Hill, A. M. van der Zande, D. A. Chenet, E.-M. Shih, J. Hone, and T. F. Heinz, Measurement of the optical dielectric function of monolayer transition-metal dichalcogenides: $MoS_2$, $MoSe_2$, $WS_2$, and $WSe2$, Phys. Rev. B **90**, 205422 (2014).

[19] R. Coehoorn, C. Haas, and R. A. De Groot, Electronic structure of $MoSe_2$, $MoS_2$, and $WSe_2$. II. The nature of the optical band gaps, Phys. Rev. B **35**, 6203 (1987).

[20] E. Cappelluti, R. Roldán, J. A. Silva-Guillén, P. Ordejón, and F. Guinea, Tight-binding model and direct-gap/indirect-gap transition in single-layer and multilayer $MoS_2$, Phys. Rev. B **88**, 1 (2013).

[21] C.-H. Chang, X. Fan, S.-H. Lin, and J.-L. Kuo, Orbital analysis of electronic structure and phonon dispersion in $MoS_2$, $MoSe_2$, $WS_2$, and $WSe_2$ monolayers under strain, Phys. Rev. B **88**, 195420 (2013).

[22] J. Zeng, H. Liu, H. Jiang, Q.-F. Sun, and X. C. Xie, Multiorbital model reveals a second-order topological insulator in $1H$ transition metal dichalcogenides, Phys. Rev. B **104**, L161108 (2021).

[23] J. Jung and Y.-H. Kim, Hidden breathing kagome topology in hexagonal transition metal dichalcogenides, Phys. Rev. B **105**, 085138 (2022).

[24] W. P. Su, J. R. Schrieffer, and A. J. Heeger, Solitons in Polyacetylene, Phys. Rev. Lett. **42**, 1698 (1979).

[25] M. Ezawa, Higher-Order Topological Insulators and Semimetals on the Breathing Kagome and Pyrochlore Lattices, Phys. Rev. Lett. **120**, 026801 (2018).

[26] G. van Miert and C. Ortix, On the topological immunity of corner states in two-dimensional crystalline insulators, npj Quantum Mater. **5**, 63 (2020).

[27] M. A. J. Herrera, S. N. Kempkes, M. B. de Paz, A. García-Etxarri, I. Swart, C. M. Smith, and D. Bercioux, Corner modes of the breathing kagome lattice: Origin and robustness, Phys. Rev. B **105**, 085411 (2022).

[28] C. K. Geschner, A. Y. Chaou, V. Dwivedi, and P. W. Brouwer, On the band topology of the breathing kagome lattice, arXiv:2412.20460.

[29] B. Bradlyn, L. Elcoro, J. Cano, M. G. Vergniory, Z. Wang, C. Felser, M. I. Aroyo, and B. A. Bernevig, Topological quantum chemistry, Nature **547**, 298 (2017).

[30] J. Sødequist, U. Petralanda, and T. Olsen, Abundance of second order topology in $C_3$ symmetric two-dimensional insulators, 2D Mater. **10**, 015009 (2023).

[31] M. Holbrook, J. Ingham, D. Kaplan, L. Holtzman, B. Bierman, N. Olson, L. Nashabeh, S. Liu, X. Zhu, D. Rhodes, K. Barmak, J. Hone, R. Queiroz, and A. Pasupathy, Real-space imaging of the band topology of transition metal dichalcogenides, arXiv:2412.02813.

[32] P. Törmä, Essay: Where Can Quantum Geometry Lead Us, Phys. Rev. Lett. **131**, 240001 (2023).

[33] Y. Onishi and L. Fu, Fundamental Bound on Topological Gap, Phys. Rev. X **14**, 11052 (2024).

[34] B. Ghosh, Y. Onishi, S.-Y. Xu, H. Lin, L. Fu, and A. Bansil, Probing quantum geometry through optical conductivity and magnetic circular dichroism, Sci. Adv. **10**, eado1761 (2024).

[35] M. Ezawa, Analytic approach to quantum metric and optical conductivity in Dirac models with parabolic mass in arbitrary dimensions, Phys. Rev. B **110**, 195437 (2024).

[36] G. Kresse and J. Furthmüller, Efficient iterative schemes for *ab initio* total-energy calculations using a plane-wave basis set, Phys. Rev. B **54**, 11169 (1996).

[37] G. Kresse and D. Joubert, From ultrasoft pseudopotentials to the projector augmented-wave method, Phys. Rev. B **59**, 1758 (1999).

[38] J. P. Perdew, K. Burke, and M. Ernzerhof, Generalized Gradient Approximation Made Simple, Phys. Rev. Lett. **77**, 3865 (1996).

[39] G. Pizzi, V. Vitale, R. Arita, S. Blügel, F. Freimuth, G. Géranton, M. Gibertini, D. Gresch, C. Johnson, T. Koretsune, J. Ibañez-Azpiroz, H. Lee, J.-M. Lihm, D. Marchand, A. Marrazzo, Y. Mokrousov, J. I. Mustafa, Y. Nohara, Y. Nomura, L. Paulatto, S. Poncé, T. Ponweiser, J. Qiao, F. Thöle, S. S. Tsirkin, M. Wierzbowska, N. Marzari, D. Vanderbilt, I. Souza, A. A. Mostofi, and J. R. Yates, Wannier90 as a community code: new features and applications, J. Phys: Condens. Matter **32**, 165902 (2020).

[40] G.-B. Liu, W.-Y. Shan, Y. Yao, W. Yao, and D. Xiao, Three-band tight-binding model for monolayers of group-VIB transition metal dichalcogenides, Phys. Rev. B **88**, 085433 (2013).

[41] K. Momma and F. Izumi, VESTA 3 for three-dimensional visualization of crystal, volumetric and morphology data, J. Appl. Crystallogr. **44**, 1272 (2011).

[42] R. Kubo, A general expression for the conductivity tensor, Can. J. Phys. **34**, 1274 (1956).

[43] D. A. Greenwood, The Boltzmann Equation in the Theory of Electrical Conduction in Metals, Proc. Phys. Soc. **71**, 585 (1958).

[44] See Supplemental Material for details about breathing kagome structure of h-TMDs, tight-binding models, Wannier function and Wannier center analysis, optical absorption of h-TMDs, and model analysis.

[45] N. A. Spaldin, A beginner's guide to the modern theory of polarization, J. Solid State Chem. **195**, 2 (2012).

# SUPPLEMNETAL MATERIAL

## Supplemental Material for

## Obstruction-Driven Parity Inversion for Enhanced Optical Absorption in Hexagonal Transition Metal Dichalcogenides

**Authors:** Seungil Baek[1], Jun Jung[1], and Yong-Hyun Kim[1,2,*]

[1]Department of Physics, Korea Advanced Institute of Science and Technology (KAIST), Daejeon 34141, Republic of Korea

[2]School of Physics, Institute of Science, Suranaree University of Technology, Nakhon Ratchasima 30000, Thailand

*Correspondence to: yong.hyun.kim@kaist.ac.kr

## I. Breathing kagome structure in h-TMDs

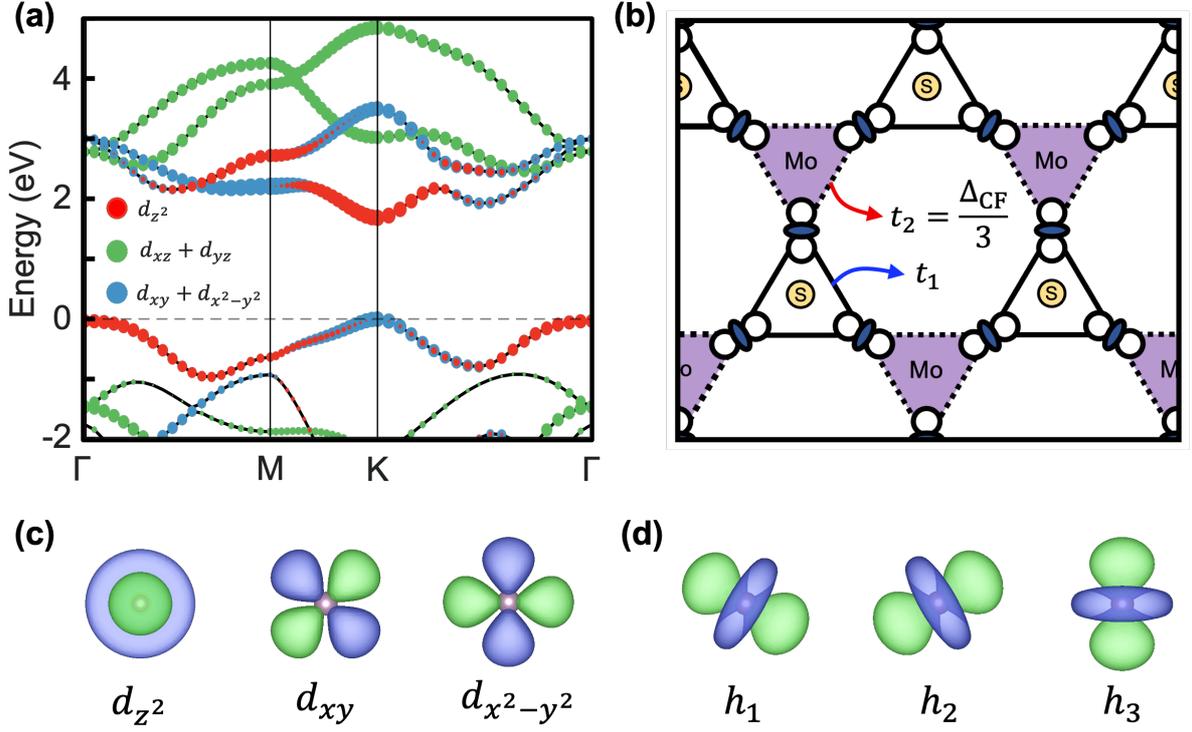

FIG. S1. Electronic structures of h-MoS$_2$. (a) Projected band structure of h-MoS$_2$. The uppermost valence band and the two lowest conduction bands are mainly composed of $d_{z^2}$, $d_{xy}$, and $d_{x^2-y^2}$ orbitals. (b) Schematic illustration of hybrid $d$ orbitals in h-TMDs. The crystal-field splitting in the atomic orbital basis transforms into on-site hopping $t_2$ between the hybrid $d$ orbitals. Intersite hopping ($t_1$) and on-site hopping ($t_2$) between hybrid $d$ orbitals result in a breathing kagome configuration. (c) Wave functions of the $d_{z^2}$, $d_{xy}$, and $d_{x^2-y^2}$ orbitals. (d) Hybrid $d$ orbitals $h_1$, $h_2$, and $h_3$, which are transformed from atomic orbitals in (a).

# SUPPLEMNETAL MATERIAL

## II. Tight-binding models

The three-band tight-binding (TB) Hamiltonians for h-TMDs are constructed in the basis of hybrid $d$ orbitals $\{h_1, h_2, h_3\}$. In momentum space, it reads

$$H(\mathbf{k}) = \sum_{\mathbf{R}} t(\mathbf{R}) e^{-i\mathbf{k}\cdot\mathbf{R}}, \qquad (S1)$$

where $t(\mathbf{R})$ denotes the hopping matrix associated with the lattice vector $\mathbf{R}$. We include up to third-nearest-neighbor hoppings. The $C_3$ rotational symmetry constrains nearest-neighbor ($h_i$), next-nearest-neighbor ($r_i$), and third-nearest-neighbor ($l_i$) hopping matrices into the following forms:

$$t(\mathbf{R}_1) = \begin{pmatrix} h_0 & h_1 & h_2 \\ h_2 & h_{11} & h_{12} \\ h_1 & h_{21} & h_{11} \end{pmatrix} \qquad (S2)$$

$$t(\mathbf{R}_1 + \mathbf{R}_2) = \begin{pmatrix} r_0 & r_1 & r_2 \\ r_1 & r_{11} & r_1 \\ r_2 & r_1 & r_0 \end{pmatrix} \qquad (S3)$$

$$t(2\mathbf{R}_1) = \begin{pmatrix} l_0 & l_1 & l_2 \\ l_2 & l_{11} & l_{12} \\ l_1 & l_{21} & l_{11} \end{pmatrix} \qquad (S4)$$

The on-site term is given by

$$t(\mathbf{0}) = \begin{pmatrix} 0 & -\frac{1}{3}\Delta_{CF} & -\frac{1}{3}\Delta_{CF} \\ -\frac{1}{3}\Delta_{CF} & 0 & -\frac{1}{3}\Delta_{CF} \\ -\frac{1}{3}\Delta_{CF} & -\frac{1}{3}\Delta_{CF} & 0 \end{pmatrix}, \qquad (S5)$$

where the $\Delta_{CF}$ denotes the crystal-field splitting between the $d_{z^2}$ orbital and $d_{xy}$, $d_{x^2-y^2}$ orbitals. The other hopping matrices could be obtained using symmetry relation and reciprocal relation $t(-\mathbf{R}) = t^T(\mathbf{R})$. All hopping parameters $h_i$, $r_i$, $l_i$ are expressed in eV and listed in Tables. S1-S3.

# SUPPLEMNETAL MATERIAL

| TB parameters (eV) | | $\Delta_{CF}$ | $h_0$ | $h_1$ | $h_2$ | $h_{11}$ | $h_{12}$ | $h_{21}$ |
|---|---|---|---|---|---|---|---|---|
| MoS$_2$ | OAL | 1.245 | 0.248 | 0.050 | 0.145 | −0.116 | **−0.822** | 0.149 |
|  | AL | **2.668** | −0.038 | 0.054 | −0.034 | 0.027 | −0.240 | 0.219 |
| MoSe$_2$ | OAL | 0.921 | 0.221 | 0.044 | 0.002 | −0.074 | **−0.777** | 0.093 |
|  | AL | **2.363** | −0.077 | −0.011 | −0.059 | 0.075 | −0.128 | 0.172 |
| MoTe$_2$ | OAL | 0.590 | 0.189 | 0.015 | −0.016 | 0.003 | **−0.677** | 0.046 |
|  | AL | **1.903** | −0.020 | −0.062 | −0.077 | 0.107 | −0.112 | 0.090 |
| WS$_2$ | OAL | 1.508 | 0.281 | 0.079 | 0.042 | −0.186 | **−0.903** | 0.177 |
|  | AL | **2.962** | −0.041 | 0.120 | −0.065 | −0.025 | −0.290 | 0.389 |
| WSe$_2$ | OAL | 1.098 | 0.244 | 0.076 | 0.023 | −0.130 | **−0.857** | 0.111 |
|  | AL | **2.697** | −0.084 | 0.012 | −0.045 | 0.034 | −0.139 | 0.223 |
| WTe$_2$ | OAL | 0.633 | 0.191 | 0.050 | 0.001 | −0.029 | **−0.745** | 0.056 |
|  | AL | **2.117** | −0.019 | −0.038 | −0.067 | 0.076 | −0.143 | 0.131 |

Table S1. Nearest-neighbor TB parameters of h-TMDs in eV. $\Delta_{CF}$ denotes the crystal-field splitting, and $h_i$ are the nearest-neighbor hoppings.

# SUPPLEMNETAL MATERIAL

| TB parameters (eV) | | $r_0$ | $r_1$ | $r_2$ | $r_{11}$ |
|---|---|---|---|---|---|
| MoS$_2$ | OAL | 0.077 | −0.005 | 0.040 | 0.048 |
| | AL | 0.077 | 0.013 | -0.034 | 0.048 |
| MoSe$_2$ | OAL | 0.069 | −0.027 | 0.015 | 0.078 |
| | AL | 0.087 | 0.004 | −0.042 | 0.043 |
| MoTe$_2$ | OAL | 0.048 | −0.059 | −0.015 | 0.122 |
| | AL | 0.079 | −0.009 | −0.067 | 0.060 |
| WS$_2$ | OAL | 0.091 | 0.013 | 0.033 | 0.019 |
| | AL | 0.080 | −0.026 | 0.050 | 0.042 |
| WSe$_2$ | OAL | 0.077 | −0.015 | 0.005 | 0.061 |
| | AL | 0.095 | 0.012 | −0.035 | 0.026 |
| WTe$_2$ | OAL | 0.045 | −0.056 | −0.033 | 0.127 |
| | AL | 0.078 | −0.001 | −0.064 | 0.062 |

Table S2. Next-nearest-neighbor hopping parameters ($r_i$) of h-TMDs in eV.

| TB parameters (eV) | | $l_0$ | $l_1$ | $l_2$ | $l_{11}$ | $l_{12}$ | $l_{21}$ |
|---|---|---|---|---|---|---|---|
| MoS$_2$ | OAL | −0.047 | 0.026 | −0.043 | 0.064 | 0.030 | −0.103 |
| | AL | −0.080 | −0.021 | −0.059 | 0.080 | 0.153 | −0.097 |
| MoSe$_2$ | OAL | −0.054 | 0.039 | −0.030 | 0.066 | −0.033 | −0.118 |
| | AL | −0.094 | −0.037 | −0.067 | 0.086 | 0.167 | −0.094 |
| MoTe$_2$ | OAL | −0.074 | 0.051 | −0.031 | 0.085 | −0.108 | −0.103 |
| | AL | −0.098 | −0.047 | −0.078 | 0.097 | 0.175 | −0.060 |
| WS$_2$ | OAL | −0.065 | 0.037 | −0.055 | 0.096 | −0.121 | −0.147 |
| | AL | −0.128 | 0.004 | −0.094 | 0.119 | 0.066 | −0.139 |
| WSe$_2$ | OAL | −0.070 | 0.047 | −0.041 | 0.086 | −0.039 | −0.146 |
| | AL | −0.123 | −0.048 | −0.085 | 0.113 | 0.177 | −0.119 |
| WTe$_2$ | OAL | −0.092 | 0.056 | −0.044 | 0.102 | −0.099 | −0.120 |
| | AL | −0.126 | −0.055 | −0.097 | 0.119 | 0.189 | −0.078 |

Table S3. Third-nearest-neighbor hopping parameters ($l_i$) of h-TMDs in eV.

**III. Wannier function analysis**

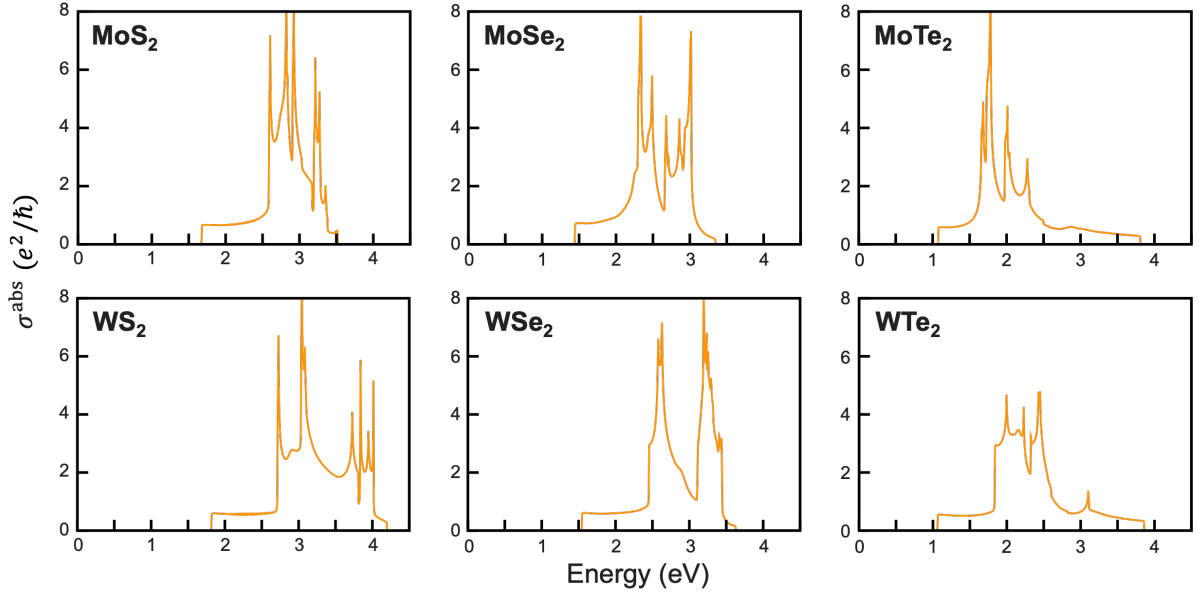

FIG. S2. Optical conductivities of h-TMDs obtained from three-band MLWF-interpolated Hamiltonians. The low-energy behavior closely matches the optical conductivities calculated from three-band TB models.

Optical conductivities ($\sigma^{abs}$) are obtained from three-band MLWF-interpolated Hamiltonians. The target bands are the same as the TB model. Calculation parameters, including *k*-points sampling and Lorentzian broadening, were identical to the TB model calculations. The general features are mainly alike, where a slight difference comes from the discrepancy between MLWF and the least-squares fitting procedure.

# SUPPLEMNETAL MATERIAL

## IV. Wannier centers of the three-band model

|  |  | Wannier center |  |  | Wannier center |
|---|---|---|---|---|---|
| MoS$_2$ | OAL | (0.666, 0.332) | WS$_2$ | OAL | (0.666, 0.333) |
|  | AL | (0.004, 0.009) |  | AL | (0.007, 0.014) |
| MoSe$_2$ | OAL | (0.667, 0.333) | WSe$_2$ | OAL | (0.667, 0.333) |
|  | AL | (0.006, 0.012) |  | AL | (0.007, 0.014) |
| MoTe$_2$ | OAL | (0.666, 0.333) | WTe$_2$ | OAL | (0.668, 0.335) |
|  | AL | (0.002, 0.004) |  | AL | (0.010, 0.020) |

Table S4. Wannier centers of the OAL and AL phases of h-TMDs in fractional coordinates.

Wannier center refers to the position of the Wannier function. For a particular band index $n$, the definition of the Wannier center $\bar{\mathbf{r}}_n$ is

$$\bar{\mathbf{r}}_n = \int d^3\mathbf{r}\, \omega_n^*(\mathbf{r})\mathbf{r}\omega_n(\mathbf{r}) \tag{S6}$$

where $\omega_n(\mathbf{r})$ is the Wannier function of the $n$-th band.

In the periodic boundary condition, the Wannier center could be represented in terms of momentum space as follows.

$$\bar{\mathbf{r}}_n = i\, \frac{\Omega}{(2\pi)^3} \int_{BZ} d^3\mathbf{k}\, e^{-i\mathbf{k}\cdot\mathbf{R}} \left\langle u_{n\mathbf{k}} \left| \frac{\partial}{\partial \mathbf{k}} u_{n\mathbf{k}} \right. \right\rangle \tag{S7}$$

Here, $\Omega$ is the cell volume, and $u_{n\mathbf{k}}$ is the periodic part of the Bloch function of band index $n$ and at $\mathbf{k}$ point [52].

## V. Extension to h-TMDs

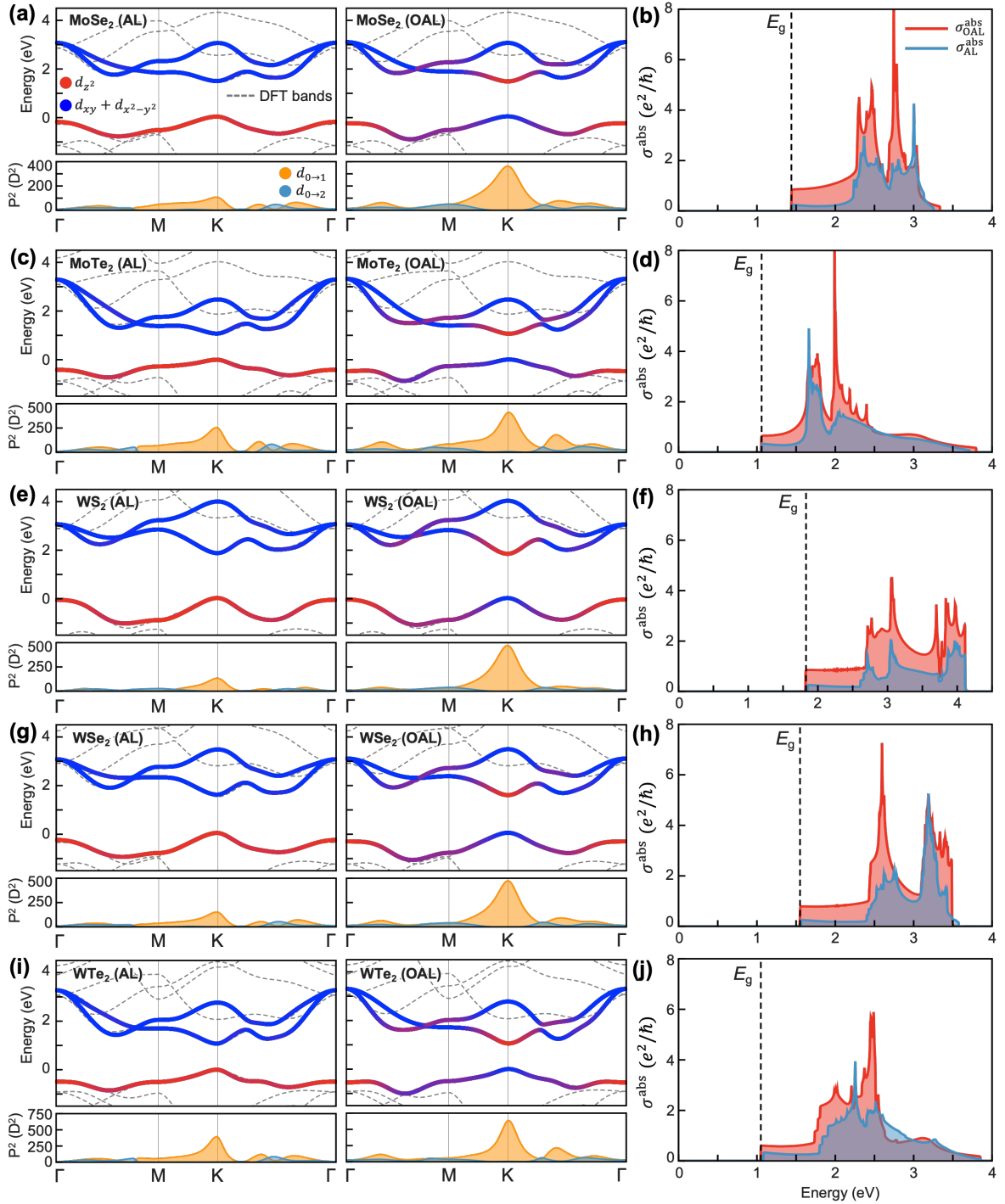

FIG. S3. Comparison of transition dipole moments and optical conductivities of the AL and OAL phases of h-TMDs. (a, b) $MoSe_2$, (c, d) $MoTe_2$, (e, f) $WS_2$, (g, h) $WSe_2$, (i, j) $WTe_2$. In all cases, the OAL phase exhibits a higher maximum transition dipole moment and optical conductivities.

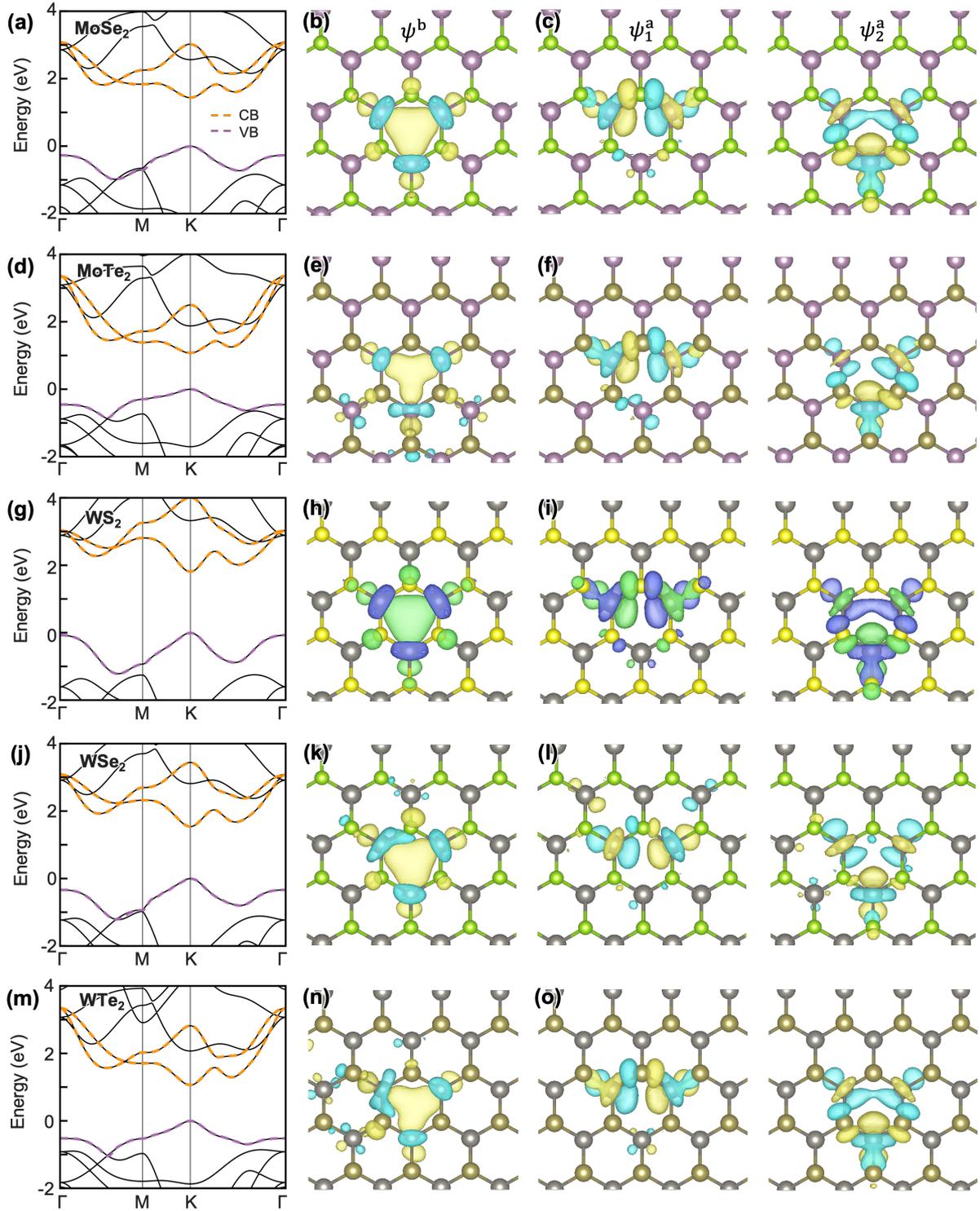

FIG. S4. Maximally localized Wannier functions (MLWFs) for h-TMDs. Wannier-interpolated band structures are compared with DFT band structures, and the MLWFs of the VBM and CB are illustrated sequentially for (a–c) $MoSe_2$, (d–f) $MoTe_2$, (g–i) $WS_2$, (j–l) $WSe_2$, and (m–o)

# SUPPLEMNETAL MATERIAL

WTe$_2$. As in MoS$_2$, the MLWFs of the valence bands correspond to trimer bonding states, whereas those of the conduction bands correspond to trimer antibonding states.

## VI. Model analysis

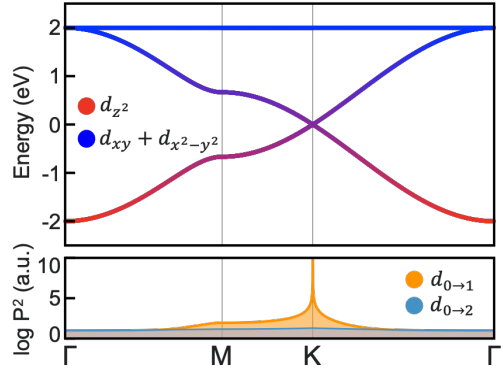

FIG. S5. Electronic band structure and log-scale transition dipole moment of simplified model systems in the exact kagome phase ($t_1 = -2/3$ eV and $t_2 = 2/3$ eV). The absence of a band gap at the K point leads to the divergence of the transition dipole moment.

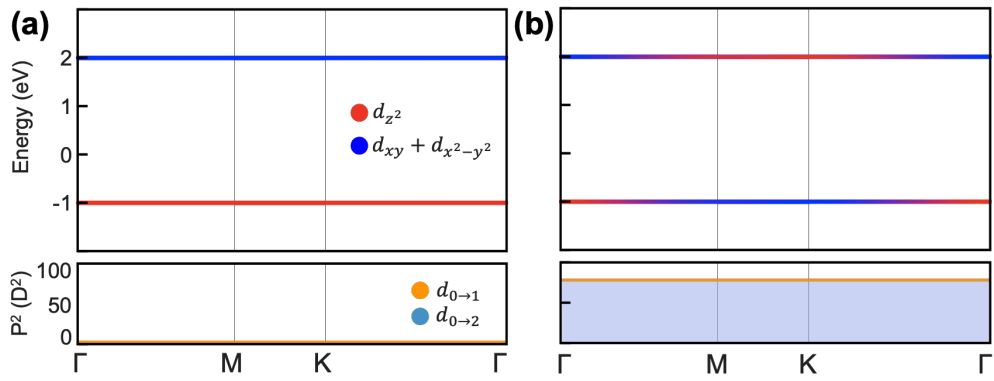

FIG. S6. Comparison of single-parameter simplified model systems in the crystal-field and hybridization limits. (a) Crystal-field-only limit ($t_1 = 0$ eV and $t_2 = 1$ eV). (b) Intersite-hopping-only limit ($t_1 = -1$ eV and $t_2 = 0$ eV). Note that band inversion significantly enhances optical transitions due to the difference in orbital character.

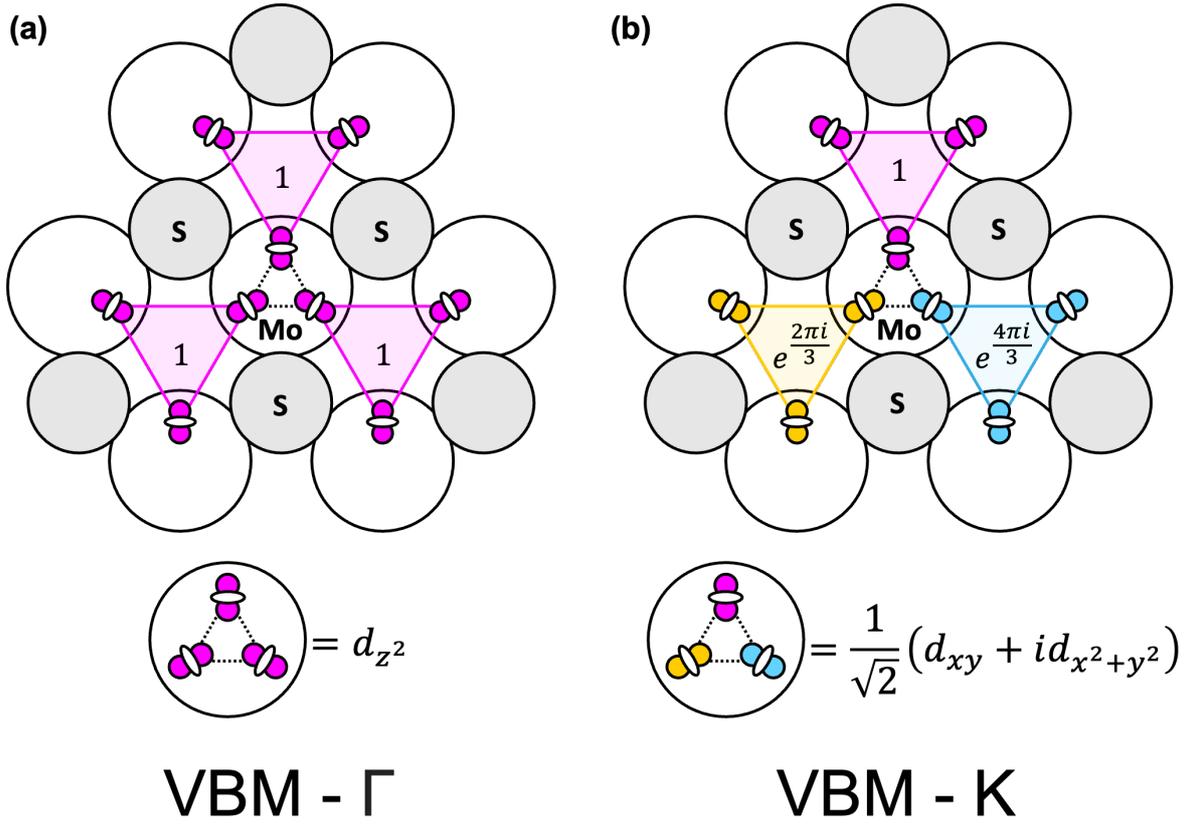

FIG. S7. Orbital superposition mechanism for band inversion at the K point. (a) An identical phase of three hybrid $d$ orbitals results in $d_{z^2}$ orbital at the VBM $\Gamma$ point. (b) A relative phase difference of $e^{2\pi i/3}$ at the K point results in a mixture of $d_{xy}$ and $d_{x^2-y^2}$ orbitals at the VBM K point.